\documentclass[preprint,3p,twocolumn]{elsarticle}
\usepackage{lineno,hyperref}
\modulolinenumbers[5]

\usepackage{url}
\usepackage{multirow}
\usepackage{amsmath,amssymb,amsfonts}
\usepackage{diagbox}
\usepackage{algorithmic}
\usepackage{graphicx}
\usepackage{textcomp}
\usepackage{xcolor}
\def\BibTeX{{\rm B\kern-.05em{\sc i\kern-.025em b}\kern-.08em
    T\kern-.1667em\lower.7ex\hbox{E}\kern-.125emX}}

\usepackage{tikz}
\usepackage{xspace}
\usepackage{algorithm2e}
\usepackage{caption}
\usepackage{subcaption}
\usepackage{url}

\usepackage{listings}
\newcommand{\mm}[1]{{\small{\textsf{#1}}}}
\newcommand{\etal}[0]{\emph{et al.}\xspace}

\newcommand\ie{\emph{i.e.},\xspace}
\newcommand\eg{\emph{e.g.},\xspace}
\newcommand{\citesec}[1]{Section~\ref{sec:#1}}
\newcommand{\citefig}[1]{Figure~\ref{fig:#1}}

\newcommand{\citetable}[1]{Table~\ref{table:#1}}

\def\Dnnf{{\tt DNNF}}
\def\dDnnf{{\tt d-DNNF}}
\def\cnf{{\tt CNF}}
\def\dnf{{\tt DNF}}
\def\obdd{{\tt OBDD}}

\def\mddg{{\tt MDDG}}

\newcommand{\eat}[1]{}

\newcommand{\cachet}{\mm{cachet}}
\newcommand{\dfour}{\mm{d4}}
\newcommand{\cdd}{\mm{c2d}}
\newcommand{\sharpsat}{\mm{sharpSAT}}
\newcommand{\Dsharp}{\mm{DSharp}}
\newcommand{\maxsat}{\mm{MaxSAT}}
\newcommand{\uwrmaxsat}{\mm{UWrMaxSAT}}
\newcommand{\maxhs}{\mm{MaxHS}}
\newcommand{\satfourj}{\mm{Sat4j}}

\newcommand{\softName}{\mm{Winston}}

\usepackage{changepage}
\usepackage{framed}
\makeatletter
 {\par\unskip\endMakeFramed}
\makeatother

\definecolor{formalshade}{rgb}{0.93,0.93,0.93}
\definecolor{darkblue}{rgb}{0.2, 0.2, 0.2}

\newenvironment{formal}{%
  \def\FrameCommand{%
    \hspace{1pt}%
    {\color{darkblue}\vrule width 2pt}%
    {\color{formalshade}\vrule width 4pt}%
    \colorbox{formalshade}%
  }%
  \MakeFramed{\advance\hsize-\width\FrameRestore}%
  \noindent\hspace{-1pt}% disable indenting first paragraph
  \begin{adjustwidth}{}{7pt}%
  \vspace{2pt}\vspace{2pt}%
}
{%
  \vspace{3pt}\end{adjustwidth}\endMakeFramed%
}

\newcounter{resultcounter}

\begin{document}

\begin{frontmatter}

\title{Reasoning on Feature Models:\\ Compilation-Based vs. Direct Approaches}

%\author{Pierre Bourhis, Laurence Duchien, Jérémie Dusart, Clément Quinton}
%\address{}
\author[lille]{Pierre Bourhis}
\author[lille]{Laurence Duchien}
\author[lille]{J\'er\'emie Dusart}
\author[lens]{Emmanuel Lonca}
\author[lens,IUF]{Pierre Marquis}
\author[lille]{Cl\'ement Quinton}

\address[lille]{University of Lille, CNRS, Inria, Centrale Lille, UMR 9189 CRIStAL}
\address[lens]{Univ. Artois, CNRS, UMR 8188 CRIL}
\address[IUF]{Institut Universitaire de France}

\begin{abstract}
Analyzing a Feature Model (FM) and reasoning on the corresponding configuration space is a central task in Software Product Line (SPL) engineering. 
Problems such as deciding the satisfiability of the FM and eliminating inconsistent parts of the FM have been well resolved by translating 
the FM into a conjunctive normal form (\cnf) formula, and then feeding the \cnf\ to a SAT solver. 
However, this approach has some limits for other important reasoning issues about the FM, such as counting or enumerating configurations. 
Two mainstream approaches have been investigated in this direction: (i) direct approaches, using tools based on the \cnf\ representation of the FM at hand,
or (ii) compilation-based approaches, where the \cnf\ representation of the FM has first been translated into another representation for which the reasoning queries are easier to address. 
Our contribution is twofold.
First, we evaluate how both approaches compare when dealing with common reasoning operations on FM, namely counting configurations, pointing out one or several configurations, sampling configurations, and finding optimal configurations regarding a utility function.
Our experimental results show that the compilation-based is efficient enough to possibly compete with the direct approaches and that the cost of translation (\ie the compilation time) can be balanced when addressing sufficiently many complex reasoning operations on large configuration spaces.
Second, we provide a Java-based automated reasoner that supports these operations for both approaches, thus eliminating the burden of selecting the appropriate tool and approach depending on the operation one wants to perform.
\end{abstract}

\begin{keyword}
Feature Model, Configuration Space, Reasoning, Solver, Knowledge Compilation.
%\MSC[2010] 00-01\sep  99-00
\end{keyword}

\end{frontmatter}

%\linenumbers

\section{Introduction}
\label{sec:introduction}

With the emergence of highly-variable software-intensive systems such as internet-of things,  cyber-physical systems or cloud-based ones, developers now have to maintain not only a single system, but numerous variants (\ie \emph{configurations}) of that system.
That is, they have to develop, test and maintain a significant number of \emph{features} that are then combined together to produce a specific software configuration.
As the number of feature grows, the number of configurations (\ie the \emph{configuration space}) consequently grows exponentially.
Manually dealing with those configurations is not feasible, especially when the system exhibits thousands of configurations, and requires an automated tool support~\cite{Benavides_JIS_10}.

Performing automated analyses to understand and validate the configuration space is a central task in \emph{Software Product Lines} (SPL) engineering and is crucial to guarantee a proper derivation of the expected software variants. 
In particular, numerous works have focused on the automated analysis of \emph{Feature Models} (FM), a well-known approach for encoding the configuration space of configurable software systems.
These analyses are of particular importance when configuring, deriving and testing configurations of the SPL~\cite{Benavides_JIS_10} as they provide a foundation for analyzing the code of a SPL later~\cite{Apel_FOSPL_16}.
Classical automated analyses include, among others, counting or enumerating configurations.

However, this automated analysis of FM is challenging and comes at a cost in time and space. 
That is, some reasoning operations tend to have longer runtimes and even become infeasible on larger FM~\cite{DPohlLP11}.
The complexity of an analysis depends on the chosen FM representation.
More precisely, a given representation may be more suitable than other ones regarding a certain analysis.
For instance, checking the satisfiability of an \emph{Ordered Binary Decision Diagram} (\obdd) \cite{Bryant86} is an operation with constant effort~\cite{Thum20},
while there is no polynomial-time algorithm for addressing the same query when the FM is represented as a \cnf\ formula (and it is likely that
no such algorithm may exist, since it would show that {\sf P = NP}).
Yet, \obdd\ has proved not to scale well when dealing with large configuration spaces such as the one of Linux~\cite{Thum20} or the ones related to highly-configurable systems~\cite{Sundermann_VAMOS_20}.
To perform automated analysis on FM, one thus has to \emph{(i)} select the appropriate FM representation and \emph{(ii)} find a tool that properly handles such representation and analysis.

To tackle these issues, we propose in this paper an empirical study that evaluates how different representations compare when addressing different types of analysis. 
In addition, we provide \softName, a Java-based library that takes as input FM representations in different formats and computes automated analysis relying on the appropriate tool support.

The remainder of the paper is structured as follows. 
Section~\ref{sec:representations} provides background information on FM and their translations into propositional representations.
Section~\ref{sec:requests} classifies automated analyses into three families.
Section~\ref{sec:direct} and Section~\ref{sec:compilation} describe the direct and compilation-based approaches for translating FM, respectively.
Section~\ref{sec:soft} provides an overview of our Java-based library, \softName.
Section~\ref{sec:method} presents our experimental methodology while Section~\ref{sec:results} discusses the empirical results.
Section~\ref{sec:conclusion} concludes the paper.

\section{Feature Model Representations}
\label{sec:representations}

A feature model is a tree or a directed acyclic graph of features~\cite{SPL-FOSE14}, organized hierarchically in parent - sub-feature(s) relationships. 
Features can be mandatory, optional or alternative, and the selection of a feature may require or exclude the selection of other ones.
While most of these relationships can be encoded in a feature tree, require and exclude relationships are usually defined as cross-tree constraints.
Therefore, the FM describes the configuration space of a software system encoded \emph{(i)~}as a feature tree and \emph{(ii)~}a set of cross-tree constraints.
It thus defines, in an implicit yet compact way, the set of possible configurations that can be derived by the SPL to yield a software product.
\citefig{example} depicts the well-known FM for mobile phone software originating from \cite{Benavides_JIS_10}.
The \mm{Calls} feature is mandatory, the \mm{GPS} feature is optional while features \mm{Screen} and \mm{Media} define \emph{alternative} and \emph{or} relationships with their sub-features, respectively.
The \mm{Camera} feature requires the \mm{High resolution} one, and features \mm{GPS} and \mm{Basic} exclude each other.
\vspace{-1.3em}

\begin{figure}[htb]
  \centering
    \includegraphics[width=\columnwidth]{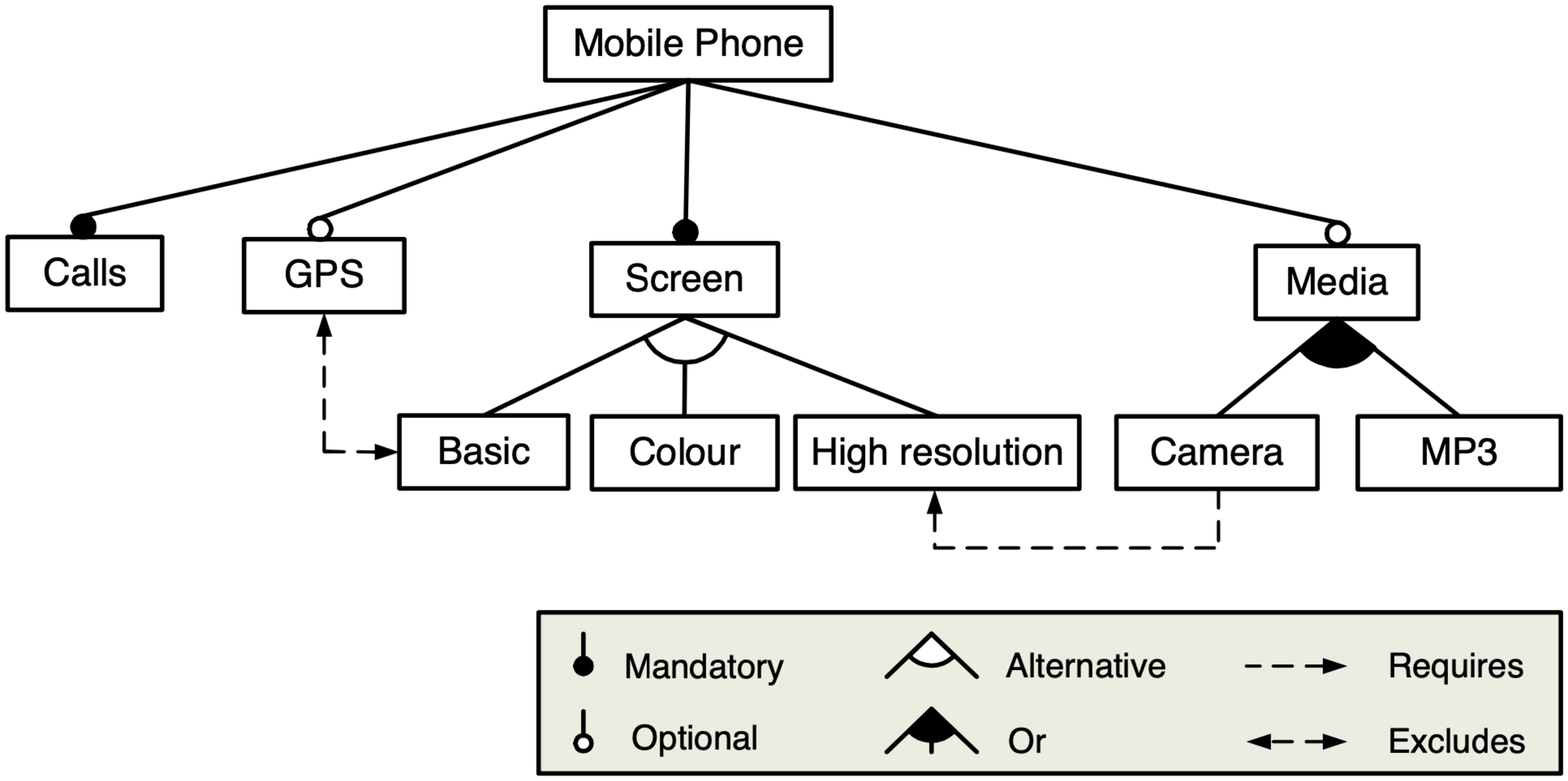}
      \vspace{-1.1em}
  \caption{Sample feature model from \cite{Benavides_JIS_10}}
  \label{fig:example}
\end{figure}

To perform reasoning operations on FM, one can take advantage of automated reasoning tools that use as input some
specific logical representations of the FM. Depending on the representation used, the computational effort is made once
or each time a reasoning operation must be conducted, as depicted in~\citefig{comparison}.
On the one hand, using the \emph{direct} approach, the FM at hand is first turned into a specific propositional representation (namely, a conjunctive normal form or \cnf\ formula) and then one takes advantage of this representation to perform reasoning queries on the FM.
For instance, \citefig{cnf:toy} shows a \cnf\ formula encoding the FM depicted in \citefig{example}.
The pros are that the translation into \cnf\ is computationally
easy (in linear time via the introduction of auxiliary variables \cite{Tseitin68,PlaistedG86}), the cons are that standard reasoning 
queries (like model counting) are usually intractable from a \cnf\ formula, and
that the computational cost must be paid each time a reasoning query is made.

\tikzstyle{langnode}=[draw, minimum height=1cm, text width=1.5cm, align=center]
\tikzstyle{langnodeL}=[draw, minimum height=1cm, text width=2cm, align=center]
\tikzstyle{qnode}=[text width=1cm, align=center]
\tikzstyle{qarrow}=[draw, dotted, ->, >=latex, thick, shorten >=1pt]
\tikzstyle{bluearrow}=[draw, ->, >=latex, blue, thin, shorten <=2pt]
\tikzstyle{redarrow}=[draw, dashed, ->, >=latex, red, very thick, shorten <=2pt]
\tikzstyle{rnode}=[text width=1cm, align=center]

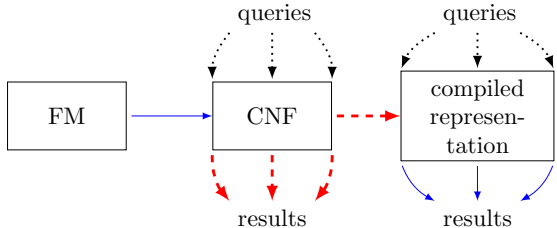
\begin{figure}[htb]
  \centering
  \scalebox{.9}{
    \begin{tikzpicture}
      \draw[white] (-1,0) -- (7,0);
      
      \node[langnode] (fm) at (0,0) {FM};
      \node[langnode] (cnf) at (3,0) {CNF};
      \node[langnodeL] (kc) at (6,0) {compiled representation};
      
      \draw[bluearrow] (fm) -- (cnf);
      \draw[redarrow] (cnf) -- (kc);

      \node[qnode] (q2) at (3,1.5) {queries};
      \node[qnode] (q3) at (6,1.5) {queries};

      \draw[qarrow] (q2.south west) to[bend right=20] (cnf.north west);
      \draw[qarrow] (q2) -- (cnf);
      \draw[qarrow] (q2.south east) to[bend left=20] (cnf.north east);
      \draw[qarrow] (q3.south west) to[bend right=20] (kc.north west);
      \draw[qarrow] (q3) -- (kc);
      \draw[qarrow] (q3.south east) to[bend left=20] (kc.north east);

      \node[rnode] (r2) at (3,-1.5) {results};
      \node[rnode] (r3) at (6,-1.5) {results};
      
      \draw[redarrow] (cnf.south west) to[bend right=20] (r2.north west);
      \draw[redarrow] (cnf) -- (r2);
      \draw[redarrow] (cnf.south east)  to[bend left=20]  (r2.north east);
      \draw[bluearrow] (kc.south west)  to[bend right=20] (r3.north west);
      \draw[bluearrow] (kc) -- (r3);
      \draw[bluearrow] (kc.south east) to[bend left=20]  (r3.north east);
      
    \end{tikzpicture}
  }
  \caption{Comparison between the direct and compilation-based approaches. The former only requires a \cnf\ representation of the FM while the latter needs a compiled representation of the \cnf\ encoding. Red dashed arrows depict expensive or intractable operations, while blue plain arrows show easy ones.}
  \label{fig:comparison}
\end{figure}

\begin{figure}[tb]
  \centering
  \[
  \begin{array}{c c c l}
    a                       & g \vee h \vee -f        &~& a: \small{\text{Mobile Phone}} \\
    b \vee -c               & a \vee -i               &~& b: \small{\text{Screen}} \\
    b \vee -d               & a \vee -j               &~& c: \small{\text{Basic}} \\
    b \vee -e               & a \vee -b               &~& d: \small{\text{Color}} \\
    c \vee d \vee e \vee -b & a \vee -f        &~& e: \small{\text{High Resolution}} \\
    -c \vee -d              & i \vee -a               &~& f: \small{\text{Media}} \\
    -c \vee -e              & b \vee -a               &~& g: \small{\text{Camera}} \\
    -d \vee -e              & -g \vee e               &~& h: \small{\text{MP3}} \\
    f \vee -g               & -j \vee -c              &~& i: \small{\text{Calls}} \\
    % f \vee -h               & -c \vee -j              &~& j: \text{GPS} \\
    f \vee -h               & ~                       &~& j: \small{\text{GPS}} \\
  \end{array}
  \]
%  \Description{A \cnf\ formula with 10 variables and 20 clauses}
  \vspace{-1.1em}
  \caption{A \cnf\ formula corresponding to the \mm{Mobile Phone} FM and the mapping between the features and the propositional variables.}
  \label{fig:cnf:toy}
\end{figure}

On the other hand, the \emph{compilation-based} approach follows a two-step process.
First, one starts with an off-line compilation step, which consists in translating the \cnf\ representation of the FM into a specific Boolean circuit, \ie a compiled representation.
Some families of compiled representations are well-known among SPL practitioners, \eg~Ordered Binary Decision Diagram (\obdd) \cite{Bryant86} or \dnf.
Second, reasoning queries are performed based on the compiled representations that result from the 
compilation step. For this approach, the cons are that the translation into a compiled form can be computationally
expensive, the pros are that this translation needs to be done only once, and that standard reasoning queries are 
tractable from those compiled representations (\eg~model counting can be achieved in linear time from an \obdd\ representation).

\section{Reasoning on Feature Models}
\label{sec:requests}
The key challenge when managing variability is to deal with the configuration space encoded by the feature model.
There is thus a huge effort in the SPL community to reason on the feature model through \emph{automated analyses}~\cite{10.1007/s00607-018-0646-1}, ranging from satisfiability checking to sampling through \eg model checking or staged configuration~\cite{Medeiros_ICSE_16, Czarnecki2005, MACHADO20141183, Classen2011}.
In this paper, we assess and compare the capabilities of the direct and compilation-based approaches on commonly used analyses.
In particular, we classify and study three \emph{families} of analyses.

\paragraph{Pointing out Configurations}
The goal here is to list some configurations whatever they are, \ie enumerating the entire set of configurations 
or only one of its proper subsets, \eg obtained by sampling the configuration space.
The problem of enumerating all configurations consists in finding all satisfiable assignments of a given propositional representation.
This problem, also known as the all-SAT (or model enumeration) problem, finds application in various domains such as model checking, automata construction or backbone computation~\cite{khurshid-04, 10.5555/1860967.1860972, jabbour-14, 10.1145/3324884.3416569}. In software engineering, enumerating configurations is necessary to test, model-check or measure performance of each derived software variants~\cite{Galindo_SPLC_16}.
However, when facing large configuration spaces, enumerating all configurations is not possible.
A well-known alternative consists in sampling the configuration space.
To make this subset as representative as possible from all configurations, a considerable amount of research has been conducted on random uniform sampling techniques applied to feature models \cite{8812049, 10.1145/3106237.3106273, 10.1145/3336294.3336297, 10.1007/978-3-319-94144-8_9, LPAR-22:Knowledge_Compilation_meets_Uniform, 8730148}.
A particular case of of the problem at hand is when a single configuration is targeted, \ie the SAT problem: one wants to determine whether there exists an assignment for which the propositional representation evaluates to true. This problem is {\sf NP}-complete in the general case; in practice, it can be solved by turning the input representation into an equisatisfiable \cnf\ formula (this can be achieved in linear time in the size of the representation via the introduction of new variables \cite{Tseitin68,DBLP:journals/jsc/PlaistedG86}), and then leveraging a SAT solver to decide whether the resulting \cnf\ formula is satisfiable or not.

\paragraph{Counting Configurations}
Counting configurations consists in computing the number of valid full assignments of the propositional representation of the FM.
Counting configurations is useful \emph{(i)~}\emph{per se} since it is easier to manage variability when the number of valid configurations is known and \emph{(ii)~}to compute various properties such as commonality, homogeneity, rating errors or variability reduction. It is also a key to random uniform sampling.
That mentioned, counting models (the \#SAT problem) is known as a computationally demanding problem (the problem is {\sf \#P}-complete).
Indeed, the \#SAT problem looks actually harder than the SAT problem: the complexity gap between the two problems is assumed from 
the theoretical side (as reflected by Toda's theorem \cite{DBLP:journals/siamcomp/Toda91} showing that {\sf PH} $\subseteq$ {\sf P}$^{\sf \#P}$, i.e., every problem
from the polynomial hierarchy {\sf PH} can be solved in polynomial time provided that a {\sf \#P} oracle is available), but it can also be observed in practice.
Thus, Sundermann \etal recently reported that some FM with approx. 1700 features could not be counted within a 24h timeout~\cite{Sundermann_VAMOS_20}. 
Especially, while counting the number of configurations of a feature tree can be done with a linear time complexity, counting the number of configurations 
of a feature model (feature tree and cross-tree constraints) cannot be performed in polynomial time.

\paragraph{Finding Optimal Configurations}
Finding optimal configurations related to functional or non-functional requirements consists in pointing out a configuration (or several configurations, or even all configurations) that minimizes or maximizes a given linear utility (or disutility/cost) function. Having the ability to address this query efficiently is very useful to handle a number of situations of interest. 
For instance, let us consider the following scenarios, where a stakeholder wants to:
\begin{description}
\item[- $S_1$:] enumerate the lowest values related to configurations, \eg enumerate the three lowest prices among all configuration prices;
\item[- $S_2$:] rank the best configurations regarding a given criteria, \eg rank the three less energy consuming configurations.
\end{description}

Both scenarios guide the user when selecting a configuration by narrowing the configuration space to a limited subset of configurations.
To address such scenarios, we rely on the top-$k$ approach proposed very recently in \cite{bourhis2022pseudo}. 
Our top-$k$ approach is based on pseudo polynomial-time algorithms\footnote{Pseudo polynomial-time means here that our algorithms run in time polynomial in
the value of $k$, and not in the size of $k$ (represented, as usual, in binary notation). Since the value of $k$ that is considered is typically small, the algorithms prove efficient
in practice.} able to \emph{(i)} rank values related to configurations, given a (dis)utility function (scenario $S_1$), hereafter referred to as the \emph{top-$k$ values} analysis; and \emph{(ii)} rank configurations given a certain criteria (scenario $S_2$), hereafter referred to as the \emph{top-$k$ configurations} analysis.

\section{The Direct Approach}
\label{sec:direct}

Unlike the compilation-based approach, the direct approach to FM analysis does not need the propositional representation of the FM to be preprocessed.
Instead, solvers of various kinds are used to address the queries of interest (pointing out configurations, counting them, finding optimal configurations) for FM analyses.

\subsection{SAT Solvers}\label{subsec:sat}
SAT solvers \cite{DBLP:series/faia/2009-185} have been widely used in SPL engineering to point out and count the number of configurations the FM exhibits \cite{DBLP:journals/ijseke/HeradioFCA13, DBLP:journals/corr/abs-1007-1024}. 

The SAT problem consists in determining whether a given propositional formula is satisfiable.
When the propositional representation of the FM at hand is satisfiable, the FM is non void since it has at least one valid configuration.

In this paper, we pointed out and counted configurations by using a SAT solver. % repetitively for enumerating models of a CNF. 
The approach is iterative and proceeds as follows: at each step, one uses a SAT solver to determine whether the current propositional representation is satisfiable
(one starts with the propositional representation of the FM); if not satisfiable, the procedure stops; otherwise, the model found is reported, and a clause equivalent to
the negation of this model is conjoined with the propositional representation at hand, to produce a new propositional representation. The procedure
then resumes. At each step, the clause that is added (a so-called blocking clause) prevents the model just found to be found at the subsequent steps.
For instance, supposing that $\{x_1,\overline{x_2},x_3,\overline{x_4}\}$ is the model found at a given step, then the clause $\overline{x_1}\vee x_2 \vee \overline{x_3} \vee x_4$ is added to the current representation and the SAT solver is called again to get another model (if any).
Obviously enough, such an enumeration-based approach to counting requires an exponential amount of time when the number of models is itself
exponential in the number of variables (in some sense, this approach to model counting relies on a base $1$ representation of the models count).
\#SAT solvers like  \cachet\footnote{\url{www.cs.rochester.edu/~kautz/Cachet/index.htm}} \cite{Sangetal04} can be exploited to avoid this drawback (they allow several models to be counted at a time). However, we did not include \#SAT solvers into our investigation because many state-of-the-art \#SAT solvers follow a compilation-based approach,
and the remaining ones are really close to compilers. Indeed, compiled representations in the \Dnnf\ language or some of its subsets correspond typically to the trace of
the computation achieved by \#SAT solver \cite{HuangDarwiche07}. For instance, the compiler 
\Dsharp\footnote{\url{https://github.com/QuMuLab/dsharp}} \cite{Muiseetal12} relies on the model counter 
\sharpsat\footnote{\url{sites.google.com/site/marcthurley/sharpsat}}\cite{Thurley06}.

\subsection{MaxSAT Solvers}

The MaxSAT problem is a generalization of the SAT problem, where given a \cnf\ formula and an integer $k$, one is interested in deciding whether at least $k$ clauses of the \cnf\ formula can be satisfied simultaneously (thus, when the input \cnf\ formula contains $m$ clauses, deciding whether it is satisfiable boils down to solving the corresponding instance of MaxSAT with $k = m$). The optimization problem associated with MaxSAT amounts to determining a subset of the input clauses that can be satisfied simultaneously and is as large as possible. The  weighted partial \maxsat\ problem further generalizes \maxsat\ by partitioning the clauses of the input \cnf\ formula into ``hard" clauses that must be satisfied, and ``soft" clauses. With each ``soft" clause, a weight representing the cost of not satisfying the clause is associated. 
The goal is then to find an assignment that satisfies all the ``hard" clauses while maximizing the sum of the weights of the ``soft" clauses that are satisfied. 
Unsurprisingly, as a generalization of SAT, the weighted partial \maxsat\ problem is {\sf NP}-hard. However, benefiting from the progress of modern SAT solvers on which they are based, \maxsat\ solvers greatly improved for the past decade. Among several other solvers, \maxhs~\cite{DBLP:conf/cp/DaviesB11,davies2013solving,10.1007/978-3-642-39071-5_13,10.1007/978-3-642-40627-0_21} and \uwrmaxsat\ \cite{9288239} look as quite efficient (see the results of the last \maxsat\ evaluation held in 2020\footnote{\url{https://maxsat-evaluations.github.io/2020/rankings.html}}).

Interestingly, weighted partial \maxsat\ solvers can be leveraged to finding optimal configurations. 
One starts with a propositional representation of the FM as a \cnf\ formula, forming the set of ``hard" clauses and one encodes the utility function as the set of ``soft" clauses. 
For a minimization operation, if for example literal $\overline{x_i}$ has cost $k$, the weight of the soft clause $x_i$ is set to $k$. For a maximization operation, if for example literal $\overline{x_i}$ has cost $k$, the weight of the soft clause $x_i$ is set to $M-k$ with $M$ being an upper bound of the maximal value taken by the cost function. 
If one wants to compute more than one optimal configurations (and specifically if one wants to compute how many they are), one can use the same enumeration-based
approach as considered for pointing out configurations using a SAT solver (\ie clauses are added to in the set of ``hard" clauses in an iterative way to block the optimal solutions already found). The main difference is that the cost of each optimal solution found must be evaluated at each step (but this is not demanding operation): when the cost changes,
the procedure must be stopped. 

\subsection{Pseudo-Boolean Solvers}

A last family of solvers that can be used for FM analysis consists of pseudo-Boolean (PB) solvers.  
Pseudo-Boolean solvers deal with propositional representations, called PB constraints, that generalize \cnf\ formulae.
A PB constraint is a conjunction of linear inequations over Boolean variables, having the form $\sum_{i=1}^{k} c_i \cdot x_i \geq c$  or 
$\sum_{i=1}^{k} c_i \cdot x_i \leq c$  where each $c_i$ ($i \in [k]$) and $c$ are integers. 
Such a linear inequation is satisfied by a truth assignment over the variables occurring it it when the inequality holds once
the variables have been replaced by $1$ or by $0$, as given by the truth assignment.
Every clause $x_1 \vee \ldots \vee x_p \vee \overline{x_{p+1}} \vee \ldots \vee  \overline{x_{p+q}}$ can be turned in linear time
into the equivalent PB constraint $\sum_{i=1}^p x_i - \sum_{i=1}^q x_{p+i} \geq 1 - q$, that is equivalent to 
$\sum_{i=1}^p x_i + \sum_{i=1}^q (1 - x_{p+i}) \geq 1$ (this reflects the fact that satisfying a clause consists precisely in satisfying at least one literal in it,
and that satisfying a negative literal in a clause amounts to set to $0$ the variable it is built upon).

Because of the translation above, the satisfiability problem for PB constraints is {\sf NP}-hard as well. 
However, the language of PB constraints is strictly more succinct than the \cnf\ language: a single pseudo-Boolean constraint can represent 
an exponential number of clauses \cite{DBLP:journals/jair/DixonGP04},  and this explains why it is useful as a knowledge representation
language \cite{DBLP:conf/ijcai/BerreMMW18}. Obviously enough, the enumeration-based approach to the generation of configurations
and their counting, based on blocking clauses, can be easily extended to PB constraints. Furthermore, the generation of
optimal configurations can also be achieved in a simple way by taking advantage of PB solvers, because linear (dis)utility functions 
(as weighted sums of literals of the form $\sum_{i= 1}^{n} w_i \cdot x_i + w'_i \cdot \overline{x_i}$) 
are in essence very close to PB constraints. The approach used, referred to as ``linear search'', is iterative: once a model of the input PB constraint 
(over $\{x_1, \ldots, x_n\}$) has been found (if any), one computes its value $w$ for the objective function and
one conjoins the current PB constraint with the linear inequation $\sum_{i= 1}^{n} w_i \cdot x_i + w'_i \cdot \overline{x_i} \geq w+1$
if a maximal solution is targeted, and with $\sum_{i= 1}^{n} w_i \cdot x_i + w'_i \cdot \overline{x_i} \leq w-1$
if ones looks for a minimal solution. Then the procedure resumes. An optimal solution is found when the resulting
PB constraint is unsatisfiable. In this paper, we used the PB solvers  \satfourj~\cite{DBLP:journals/jsat/BerreP10}.

\section{The Compilation-based Approach}
\label{sec:compilation}

\subsection{Overview}

There exist numerous work putting in practice the compilation-based approach to FM analysis.
These works mainly rely on a translation of the FM at hand into an \obdd\ circuit to perform various kinds of queries on the model: counting, slicing, model checking, etc.~\cite{MendoncaBC09,DPohlLP11,Sundermann_VAMOS_20,10.1007/978-3-642-24485-8_47}.
Unfortunately, \obdd\ has proved not to scale well when dealing with large configuration spaces such as the one of Linux \cite{Thum20} or the ones related to highly-configurable systems \cite{Sundermann_VAMOS_20}.
Despite these unappealing results, the compilation-based approach looks as well-suited for performing reasoning queries on large FM.
First, investing some time in compiling the FM into a more tractable representation is beneficial for queries that will be performed later on, as most of the time, several reasoning queries are combined during the FM analysis process.
Each of these queries is computationally easy from a compiled representation (in practice, achieving it mainly requires to browse the representation) while it is {\sf NP}-hard
for unrestricted propositional representations.
Second, \obdd\ is not the only language of compiled representations enabling tractable reasoning queries. Other knowledge compilation languages, that achieve other
time/space trade-offs, have been defined so far.
In this paper, we focus on the compilation language called \dDnnf. This language is the set of specific propositional representations called \dDnnf\  circuits.

\subsection{d-DNNF-Based Reasoning for Feature Models}
\label{sec:ddnnf}

The benefits of using a \dDnnf\ circuit as a compiled representation of the FM under consideration come from the fact that \dDnnf\ circuits achieve a good time/space trade-off \cite{Darwiche_2002}. 
Indeed, many queries of interest (especially, deciding whether a solution exists, enumerating solutions with polynomial delay, counting solutions, pointing out optimal solutions given a linear objective function, etc.) are tractable when based on a  \dDnnf\ circuit while they are intractable in the general case.
In addition, the \dDnnf\ language is typically strictly more succinct than other languages also offering those tractable queries such as \obdd\ circuits, as proved in \cite{Darwiche_2002}.
Being strictly more succinct means that some exponential space savings can be achieved by targeting the \dDnnf\ language instead of the \obdd\ one.
These properties of \dDnnf\ circuits, the various sets of tractable queries supported by \dDnnf\ circuits and the existence of ``efficient'' compilers, explain why the \dDnnf\ language, developed two decades ago for some AI purposes (especially, model-based diagnosis) has been spreading over a number of domains that go beyond 
AI; in particular, theoretical computer science, database theory, and more recently software engineering~\cite{Baital, Heradio22}.
%%%%%%%%%%%%%%%

\begin{figure}
  \centering
  \includegraphics[width=\columnwidth]{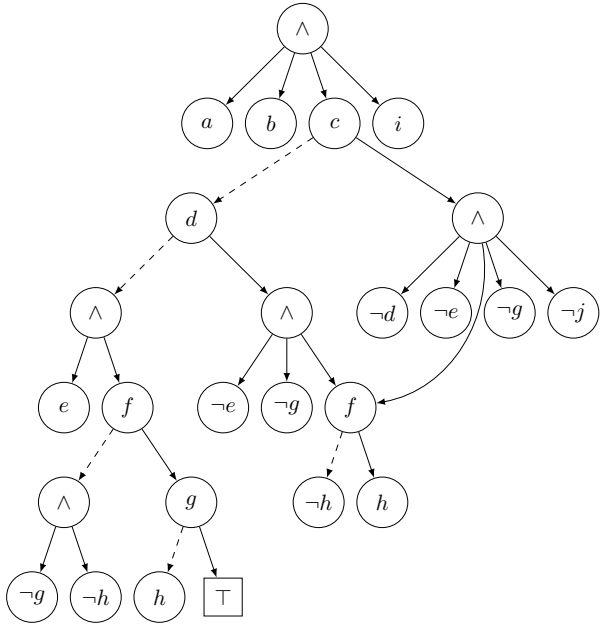}
  \caption{A Decision-\Dnnf\ circuit corresponding to the \cnf\ formula reported in \citefig{cnf:toy}.}
  \label{fig:dnnf:toy}
\end{figure}

More in detail, a \dDnnf\ circuit is a particular kind of a more general class of circuits for representing Boolean functions, called Boolean circuits. 
A Boolean circuit is a more compact representation than a Boolean formula as circuits can factorize some repeated sub-formulae.
A \emph{\dDnnf\ circuit} is a directed acyclic graph (DAG), where internal nodes are labeled by occurrences of $\wedge$ and $\vee$ connectives, 
and leaves are labeled by Boolean constants and literals. 
Any \dDnnf\  circuit respects two key properties: the sets of variables appearing in the subcircuits of the children of any $\wedge$-node are disjoint 
and the sets of models of subcircuits of the children of any $\vee$-node are pairwise disjoint.
The \dDnnf\ language includes as a proper subset the language of Decision-\Dnnf\ circuits. In a Decision-\Dnnf\ circuit, every $\vee$-node has a specific form: it has
two children, one of it being a $\wedge$-node of the form $x_i \wedge \alpha$ (where $x_i$ is a variable and 
$\alpha$ is a Decision-\Dnnf\ circuit) and its sibling is a $\wedge$-node of the form $\overline{x_i} \wedge \beta$ (where
$\beta$ also is a Decision-\Dnnf\ circuit). As a matter of illustration, a Decision-\Dnnf\ circuit corresponding to the FM example and an equivalent \cnf\ representation as
reported in \citefig{example} and~\ref{fig:cnf:toy} is provided in Figure~\ref{fig:dnnf:toy}.
What makes the Decision-\Dnnf\ language of specific interest is that top-down compilers targeting \dDnnf, like \cdd\footnote{\url{reasoning.cs.ucla.edu/c2d/}} \cite{Darwiche01,Darwiche04}, \Dsharp\footnote{\url{https://github.com/QuMuLab/dsharp}} \cite{Muiseetal12} and \dfour\footnote{\url{www.cril.univ-artois.fr/KC/d4.html}} \cite{LagniezMarquis17} actually 
targets the Decision-\Dnnf\ language.

\section{The \softName\ Reasoner}
\label{sec:soft}

To support both direct and compilation-based analysis on feature models, we implemented \softName\ (kno\textbf{W}ledge comp\textbf{I}latio\textbf{N} ba\textbf{S}ed fea\textbf{T}ure m\textbf{O}del reaso\textbf{N}er), a Java-based automated reasoner.
\softName\ is a library designed for helping the user pointing out, counting and finding optimal configurations as described in \citesec{requests}, whether the feature model is represented as a \cnf\ formula or as a \dDnnf\ circuit. 
It can be either integrated as an API to a Java application or through the scripting language detailed in \ref{sub:dsl}. 
Both \softName\ source code\footnote{\url{https://gitlab.inria.fr/jdusart/winston}} and experimental data\footnote{\url{https://gitlab.inria.fr/jdusart/kc-jss-xp}} are available online.

\subsection{Overview}

\softName\ takes as input a \cnf\ formula (in the usual DIMACS format~\cite{dimacs}) or a \dDnnf\ circuit\footnote{\url{https://github.com/crillab/d4}.}
representing the feature model to analyze. Once the propositional representation is loaded, the appropriate reasoner can be run.

The operations listed in Table \ref{table:softfunc} have been implemented in \softName\ to reason on a set of feature model configurations, hereafter referred to as the solutions to the considered problem. 
\cnf\ reasoning is based on the Java library SAT4J that provides a SAT solver, a \maxsat\ solver, and a PB solver. The SAT solver is used to point out solutions, the \maxsat\ solver to find optimal solutions, and the PB solver to enumerate optimal values. 

The input \cnf\ formula can also be compiled and transformed into a \dDnnf\ circuit using \dfour. 
The \dDnnf\ reasoning engine is implemented in a novel library that first loads the \dDnnf\ circuit generated by \dfour\, and then, depending on the requested analysis, either counts the number of solutions, samples solutions, finds top-$k$ models or finds top-$k$ values. 
\softName\ does not provide a direct approach to the counting operation since, as explained before, all the state-of-the-art \#SAT solvers compute implicitly 
or explicitly a \dDnnf\ circuit to count the number of models. Thus, comparing such a ``direct'' approach to counting with a compilation-based approach would not be very informative. For a similar reason, no ``direct'' approach to uniform sampling has been implemented in \softName.

\begin{table}[h]
\resizebox{\columnwidth}{!}{%
\begin{tabular}{ | l | c | c |  }
 \hline			
           & \cnf\        & \dDnnf\         \\
 \hline	
   Generate a set of configurations                               & \checkmark &  \checkmark     \\
   Compile to \dDnnf                                & \checkmark &    --   \\
   Count the number of configurations 		                & $\times$ &  \checkmark     \\
   Derive a set of uniformly generated configurations         & $\times$ &  \checkmark     \\
   Derive $k$ top configurations                          & \checkmark &  \checkmark     \\
   Derive the top $k$ values                          & \checkmark &  \checkmark     \\
   Achieve the top-$k$ transformation  & $\times$ &  \checkmark     \\
   \hline
 \end{tabular}
}
\caption{Operations offered by \softName\ depending on the feature model representation.}
\label{table:softfunc}
\end{table}

\subsection{The WINSTON Domain-Specific Language}
\label{sub:dsl}

\softName\ has been developed using ANTLR \cite{Parr13}, a tool for building and parsing (among others) domain-specific languages (DSL).
The \softName\ DSL is designed in such a way that function calls are similar whether the input comes as a \cnf\ or a \dDnnf, and only the loading step differs. 

Listing \ref{lst:example_cnf} illustrates the use of \softName\ on the JHipster example using the \cnf\ format.
First, the representation of the feature model is loaded as a \cnf\ formula (line 1).
Then, the analysis consists in finding an optimal configuration regarding an objective function.
In the proposed scenario, a disutility function is defined by assigning weights to features (lines 2 -- 4).
Precisely, assigning a weight of $1$ when a feature is present and of $0$ when the feature is absent 
makes \softName\ find a configuration with a minimum number of features (line 5).
Finally, the solution found (\ie \texttt{opt}) and its value are printed (lines 6 -- 7). \\

%\begin{figure}[h]
\begin{lstlisting}[caption={Finding a solution that contains as few features as possible.},captionpos=b, label={lst:example_cnf}]
rep = load_cnf("jhipster.cnf")
w = new_weighting(45) 
set_default_positive_weight(w,1)
set_default_negative_weight(w,0)
opt = optimize (rep,w,"min") 
print get_weight(w,opt) 
print opt
\end{lstlisting}
%\end{figure}

Listing \ref{lst:example_ddnnf} illustrates the use of \softName\ on the JHipster example using the \dDnnf\ representation.
First, the feature model is loaded as a \cnf\ formula and is then compiled into a \dDnnf\ circuit (line 1).
Then, a first analysis is performed as the number of solutions is counted (lines 2 -- 3).
Finally, the same analysis as the one presented in Listing \ref{lst:example_cnf} is performed.\\

\begin{lstlisting}[caption={Counting solutions and finding one that contains as few features as possible.},captionpos=b, label={lst:example_ddnnf}]
jhipster = load_cnf("jhipster.cnf")
rep = compile(jhipster)
print count(rep) 
w = new_weighting(45) 
set_default_positive_weight(w,1)
set_default_negative_weight(w,0)
opt = optimize (rep,w,"min")
print get_weight(w,opt) 
print opt 
\end{lstlisting}

\section{Experimental Methodology}
\label{sec:method}

We conducted some experiments in order to evaluate how the direct and compilation-based approaches compare and to determine the benefits that can be achieved from the practical side by leveraging a \dDnnf\ representation for analyzing feature models. In this section, we describe the experimental results we obtained regarding the three families of analyses presented in \citesec{requests}, \ie counting configurations, pointing out configurations and finding optimal configurations.

\subsection{Setup}

All the experiments have been ran on a machine from the Grid'5000 testbed~\cite{grid5000} equipped with a bi-processor Intel Xeon E5-2680 v4 (2.2 GHz) and 768 GB of memory. 
We evaluated  the different approaches on a benchmark composed of 222 features models, with the 218 feature models of \cite{10.1145/3382025.3414951} and 4 feature models (eCos, fiasco, Linux, uClinux) from \cite{10.5555/2818754.2818819}. 
This evaluation was conducted relying on several tools.
Regarding the direct approach, the CaDiCal SAT solver~\cite{CaDiCal}, considered as one of the two best SAT solver\footnote{http://fmv.jku.at/cadical/}, was used to count and point out configurations. 
The \maxhs\ MaxSAT solver and the \satfourj\ PB solver were used to compute optimal configurations, all in their default settings.
As for the compilation-based approach, the algorithm presented in~\cite{DARWICHE200481}, the KUS algorithm~\cite{SGRM18} and the Top-k algorithm from~
\cite{bourhis2022pseudo} 
were implemented in \softName\ to count, sample configurations and find optimal ones, respectively.

\subsection{Experiments}

We tried to compile each of the feature models from our dataset into an \dDnnf\ circuit by using the state-of-the-art compiler \dfour. 
Out of the 222 feature models in our dataset, only the one encoding Linux did not compile, even after a full day of computation.
All others feature models were compiled in less than 86.9s (see Figure \ref{fig:jddnnf_enum}).
In order to exploit the \dDnnf\ circuit computed by \dfour\ so as to perform the analysis of the corresponding feature model one starts with, 
this \dDnnf\ circuit must be loaded in memory. 
Indeed, \dfour\ produces as a result a text file describing the \dDnnf\ circuit that has been generated, not the circuit itself.
As we will see next, this loading phase can be time-consuming.
The time needed to compile the \cnf\ formula representing the feature model at hand and then to point out a configuration
based on the compiled form can thus be much larger than the 
time needed to solve the same problem 
when no off-line compilation of the feature model has been done upstream.
Therefore, our empirical study aims to determine in which circumstances compiling the feature model under consideration is beneficial.

\citetable{overview} provides an overview of the problems addressed and the tools used in our evaluation.
In particular, we studied how both the direct and compilation-based approaches compare for \emph{(i)} counting, \emph{(ii)} pointing out and \emph{(iii)} finding optimal configurations.
Several sub-problems were considered as pointing out problems : enumerating one solution (\ie the satisfiability problem), enumerating \emph{k} solutions or sampling configurations.
Regarding problems related to finding optimal configurations, we assessed both approaches on finding one or \emph{k} top configurations and top values, respectively.

\begin{table}[h]
\resizebox{\columnwidth}{!}{%
\begin{tabular}{llc|c|}
\hline
\multicolumn{2}{|l|}{\diagbox{\textbf{Problems}}{\textbf{Approach}}}                      & \textbf{Direct}                        	&  \textbf{d-DNNF}                                  \\ \hline
\multicolumn{2}{|l|}{\textbf{Counting configurations}}                                             & SAT+blocking clauses                        		&  implem. of  \cite{DARWICHE200481}                                   \\ \hline
\multicolumn{1}{|l|}{\multirow{3}{*}{\textbf{\shortstack{Pointing\\ out conf.}}}}   		
						      & \multicolumn{1}{l|}{$Enum_1$ (SAT)}  	 & SAT                          		&  implem. of  \cite{DARWICHE200481}                                 \\ \cline{2-4} 
\multicolumn{1}{|l|}{}                        & \multicolumn{1}{l|}{$Enum_k$}                   & SAT+blocking clauses           		&  implem. of  \cite{DARWICHE200481}                                 \\ \cline{2-4} 
\multicolumn{1}{|l|}{}                        & \multicolumn{1}{l|}{$Sampling$}                 & Feasible through counting     &  KUS \cite{SGRM18}                              \\ \hline
\multicolumn{1}{|l|}{\multirow{4}{*}{\textbf{\shortstack{Finding\\ opt. conf.\\ or values}}}}    		
						     & \multicolumn{1}{l|}{$Opt_1$}           		& MaxSAT                       		&  implem. of  \cite{bourhis2022pseudo}           \\ \cline{2-4} 
\multicolumn{1}{|l|}{}                       & \multicolumn{1}{l|}{$Opt_k$}                     	& MaxSAT+blocking clauses
               	&  implem. of  \cite{bourhis2022pseudo}  \\ \cline{2-4} 
\multicolumn{1}{|l|}{}                       & \multicolumn{1}{l|}{$Opt_{V_1}$}                & MaxSAT                       		&  implem. of  \cite{bourhis2022pseudo}          \\ \cline{2-4} 
\multicolumn{1}{|l|}{}                       & \multicolumn{1}{l|}{$Opt_{V_k}$}                	& PB solver                    		&  implem. of  \cite{bourhis2022pseudo}  \\ \hline
%\multicolumn{2}{|l|}{\textbf{Composition of problems}}                                                    & Not possible                 &  Yes                                       \\ \hline
\end{tabular}
}
\caption{Tools and algorithms used in our empirical study to address the three problems with both approaches.}
\label{table:overview}
\end{table}

In the following experiments, we compare for the direct and compilation-based approaches the \emph{success rate}, the \emph{mean time} and the \emph{max time} to perform some analysis and report them in dedicated tables. 
The success rate column gives the proportion of instances for which the goal has been reached before the timeout, the mean time column indicates the mean computation time (in seconds) required to do the job, while the rightmost column (max time) shows the largest computation time required to do the job without including the instances where it failed.

\section{Experimental Results}
\label{sec:results}

\subsection{Counting Configurations}

In this experiment, we evaluated how the direct and compilation-based approaches compare for counting the number of solutions.

\paragraph{Direct Approach}

\citetable{nonOpt_count} reports on the performance of the CaDiCal SAT solver for counting solutions with a 10 minutes timeout.
Among the 222 instances that compose the benchmark, the direct approach managed to count the number of solutions for only 12\% of them.
It is not surprising to see such a high failing rate, since our benchmark is composed of \cnf\ with a large number of solutions, and this approach never splits the problem into sub-problems but rather enumerates all the solutions. 

\begin{table}[h]
\resizebox{\columnwidth}{!}{%
\begin{tabular}{l|c|c|c|}
\cline{2-4}
                                              & success rate    & mean time  & max time  \\\hline
\multicolumn{1}{|l|}{Counting using SAT}      &  0.121          & 66.0s      & 406s     \\\hline
\end{tabular}
}
\caption{Counting solutions, direct approach.}
\label{table:nonOpt_count}
\end{table}

\vspace{-2mm}
\paragraph{Compilation-Based Approach}

We ran the experiment using the state-of-the-art compiler \dfour\ and then counted the number of solutions using \softName. 
The results are reported in \citetable{nonOpt_count_2}. % and depicted in \citefig{jddnnf_enum}. 
The first row indicates the compilation time taken by \dfour, while the second one reports on the time required by \softName\ to count solutions.
This table shows \emph{(i)} that the compilation time is significant compared to the counting time, but also \emph{(ii)} that the overall counting time remains low, especially compared to the direct approach.

\begin{table}[h]
\resizebox{\columnwidth}{!}{%
\begin{tabular}{l|c|c|c|}
\cline{2-4}
                                        & success rate    & mean time  & max time \\\hline
\multicolumn{1}{|l|}{Compilation  }     &  0.995          & 0.754 s    & 86.9 s  \\\hline
\multicolumn{1}{|l|}{Counting }         &  0.995          & 0.230 s    & 4.9 s  \\\hline
\multicolumn{1}{|l|}{Total   }          &  0.995          & 0.984 s    & 87.4 s  \\\hline
\end{tabular}
}
\caption{Counting solutions, compilation-based approach.}
\label{table:nonOpt_count_2}
\end{table}

\begin{formal}
Whenever the compilation succeeds, the counting operation is then performed much faster by the compilation-based approach.
With its low success rate, the direct approach based on a SAT solver is not competitive for counting solutions. 
\end{formal}

%\begin{figure*}[ht!]
%\centering
%\includegraphics[width=\linewidth]{pictures/compil_counting.eps}
%\caption{Time required for \dfour\ to compile each model and for \softName\ to parse the related \dDnnf\ and count solutions. }
%\label{fig:jddnnf_enum}
%\end{figure*}

\subsection{Pointing out Configurations}
In this experiment, we evaluated how the direct and compilation-based approaches compare for \emph{(i)} finding one configuration, \emph{(ii)} enumerating 10 configurations and \emph{(iii)} uniformly sampling 10 random configurations.

\paragraph{Direct Approach}

We ran the experiments consisting in finding 1 and 10 solutions for the input \cnf\ formula using the state-of-the-art CaDiCal SAT solver  with a 10 minutes timeout and reported the results in \citetable{nonOpt_CNF}. 
For 98\% of the benchmark instances, CaDiCal managed to find a solution in due time, 
It was also able to point out 10 solutions in about 3 seconds for 97\% of them. 
Our experiments thus show that the direct approach based on a SAT solver is quite efficient for pointing out solutions when the number of expected solutions is reduced.
To the best of our knowledge, there is no direct approach based on a \cnf\ representation for sampling configurations. 
Existing approaches rely on a \#SAT solver to compute probabilities and thus implicitly go through a compilation-based approach.

\begin{table}[hb!]
\resizebox{\columnwidth}{!}{%
\begin{tabular}{l|c|c|c|}
\cline{2-4}
                                                & success rate & mean time & max time  \\\hline
\multicolumn{1}{|l|}{Enum. 1 solution (SAT)}       &  0.982       &  0.01s    & 0.54 s    \\\hline
\multicolumn{1}{|l|}{Enum. 10 solutions}           &  0.973       &  0.12s    & 3.17 s    \\\hline
\multicolumn{1}{|l|}{Sampling}                  & \multicolumn{3}{c|}{Not available}    \\\hline
\end{tabular}
}
\caption{Pointing out configurations, direct approach.}
\label{table:nonOpt_CNF}
\end{table}

\paragraph{Compilation-Based Approach}

We ran the experiments consisting in finding 1 solution, finding 10 solutions and sampling 10 configurations relying on the KUS algorithm.
The KUS algorithm takes the root of a \dDnnf\ circuit as input and an integer representing the desired sample size as parameter.
It then recursively computes a uniform random sample by splitting the sampled solutions among the root children, based on the amount of solutions available for each child~\cite{SGRM18}.
The results are reported in \citetable{nonOpt_compil}.
The run times do not include the compilation time taken by \dfour, but only indicate the time required by \softName\ to load the \dDnnf\ and perform the computation.
For all benchmark instances that \dfour\ managed to compile, finding a solution to the problem was achieved in less than 5 seconds. 
Only the Linux instance compilation failed.
\vspace{5mm}

\begin{table}[h]
\resizebox{\columnwidth}{!}{%
\begin{tabular}{l|c|c|c|}
\cline{2-4}
                                              & success rate    & mean time  & max time  \\\hline
\multicolumn{1}{|l|}{Enum. 1 solution (SAT)}     &  0.995          & 0.18s      & 4.85s     \\\hline
\multicolumn{1}{|l|}{Enum. 10 solutions}         &  0.995          & 0.19s      & 4.99s     \\\hline
\multicolumn{1}{|l|}{Sample 10 solutions}        &  0.995          & 0.19s      & 4.93s     \\\hline
\end{tabular}
}
\caption{Pointing out configurations, compilation-based approach.}
\label{table:nonOpt_compil}
\end{table}

%\vspace{5mm}

\begin{formal}
The direct approach outperforms the compilation-based one for finding one solution, and also remains faster for pointing out 10 solutions.
Whenever the compilation succeeds, the compilation-based approach remains an efficient alternative for enumerating solutions.
In particular, it performs uniform sampling of 10 solutions with a similar running time as enumerating them, thus showing that while the compilation time can be important, it then pays off when performing different analysis relying on the compiled representation.
\end{formal}

\subsection{Finding Optimal Configurations}
\label{sec:topk}
In this experiment, we compare the direct and compilation-based approaches for \emph{(i)} finding $10$ top configurations and \emph{(ii)} finding the $10$ top values.

\subsubsection{Finding $10$ Top Configurations}

To assess both approaches, we considered linear (dis)utility functions.
These functions were obtained by associating with each literal of the input propositional representation an integer picked up uniformly at random in $\{0, 1, \ldots, l\}$, where $l$ is a preset bound equal to $1$, $100$, $10000$ or $1 000 000$. 
In the case of the PB solver, the functions only have weights on the variable. 
Following this approach, five value functions $\nu_1, \ldots, \nu_5$ have been generated per instance.
An instance was considered solved when the corresponding algorithm for top-$k$ solutions succeeded in deriving $k$ top solutions for each of the five value functions before the timeout was reached.
For every instance solved, we computed the mean time required to get $k$ top solutions when the value function varies.

\paragraph{Direct Approach}

Two direct approaches have been considered for the generation of $10$ top configurations, one based on the \maxhs\ partial weighted \maxsat\ solver, and one based on the \satfourj\ PB solver (in optimization mode).
The results are reported in \citetable{optConf_CNF}.
The \maxhs\ solver succeeded for every instances with only a slowdown of a factor of 2 in the worst case when the limit $l$ increased from $1$ to $100$. 
Increasing the limit beyond $100$ did not have any real impact on the performance of this solver.
The \satfourj\ solver went from a 100\% success rate when $l$ was set to $1$ to a 1\% success rate when $l$ was set to $1000000$. 
The amount of failures as well as the required computation mean time increased together with the increase of $l$.\\

\begin{table}[hb!]
\resizebox{\columnwidth}{!}{%
\begin{tabular}{l|c|c|c|}
\cline{2-4}
                                                       & success rate    & mean time  & max time  \\\hline
\multicolumn{1}{|l|}{\maxhs\ with $l=1$}           &  1.0            & 0.49s      & 26.6s     \\\hline
\multicolumn{1}{|l|}{\maxhs\ with $l=100$}         &  1.0            & 3.70s      & 55.7s     \\\hline
\multicolumn{1}{|l|}{\maxhs\ with $l=10000$}       &  1.0            & 3.92s      & 61.0s     \\\hline
\multicolumn{1}{|l|}{\maxhs\ with $l=1000000$}     &  1.0            & 3.93s      & 60.4s     \\\hline
\multicolumn{1}{|l|}{{\tt Sat4j} with $l=1$}           &  1.0            & 0.15s      & 1.16s     \\\hline
\multicolumn{1}{|l|}{{\tt Sat4j} with $l=100$}         &  0.829          & 14.3s      & 242.7s     \\\hline
\multicolumn{1}{|l|}{{\tt Sat4j} with $l=10000$}       &  0.198          & 129.5s     & 404.5s     \\\hline
\multicolumn{1}{|l|}{{\tt Sat4j} with $l=1000000$}     &  0.009          & 144.3s     & 196.2s     \\\hline
\end{tabular}
}
\caption{Top $10$ configurations, direct approach.}
\label{table:optConf_CNF}
\end{table}

\paragraph{Compilation-Based Approach}
The empirical results, reported in \citetable{optConf_dDNNF}, do not include the compilation time (shown in \citefig{jddnnf_enum}).
For each of the $221$ feature models that compiled (all but Linux), a top-$10$ configurations was derived in less than eight seconds.
As shown by the different rows, varying the $l$ bound ($1000$, $1 000 000$, $1 000 000 000$) did not impact the performance of this approach.\\

\begin{table}[h]
\resizebox{\columnwidth}{!}{%
\begin{tabular}{l|c|c|c|}
\cline{2-4}
                                                       & success rate    & mean time  & max time  \\\hline
\multicolumn{1}{|l|}{\dDnnf\ with $l=1$}                &  0.995          & 0.34s      & 7.56s     \\\hline
\multicolumn{1}{|l|}{\dDnnf\ with $l=100$}              &  0.995          & 0.34s      & 7.84s     \\\hline
\multicolumn{1}{|l|}{\dDnnf\ with $l=10000$}            &  0.995          & 0.34s      & 7.83s     \\\hline
\multicolumn{1}{|l|}{\dDnnf\ with $l=1000000$}          &  0.995          & 0.34s      & 7.96s     \\\hline
\end{tabular}
}
\caption{Top $10$ configurations, compilation-based approach.}
\label{table:optConf_dDNNF}
\end{table}

\citefig{topconf_maxhs_vs_jddnnf} compares the direct approach (using \maxhs) with the complete compilation-based approach, \ie compilation time included with $l$ set to $1000000$.
In most cases, the compilation-based approach is faster than the direct one and the compilation time pays off with one top-$10$ configurations.

\begin{figure}[ht!]
\includegraphics[width=0.9\linewidth]{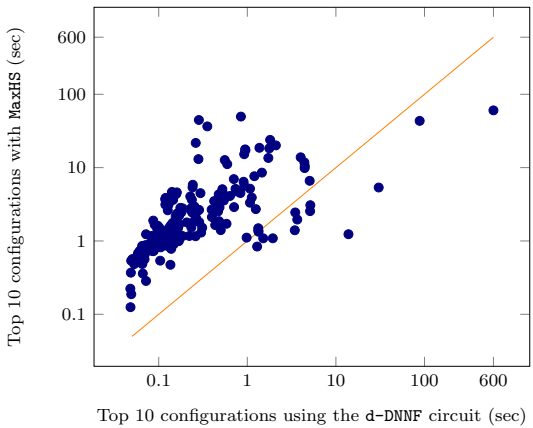}
\caption{Comparison between \maxhs\ and the compilation-based approach with compilation time included for finding top-10 configurations, with $l$ set to $1000000$.}
\label{fig:topconf_maxhs_vs_jddnnf}
\end{figure}

\begin{formal}
While the compilation time can be twice as long as the solving time in the worst-case scenario, 
the compilation-based approach outperforms on average the direct approach for generating top-$10$ configurations.
In particular, the compilation-based approach appeared empirically as ten times faster than the fastest direct approach, \ie the \maxsat\ one, and presents no sensibility to the $l$ limit. 
\end{formal}

\subsubsection{Finding the $10$ Top Values}

We performed an experiment similar to the one presented above (same instances, same timeout) but focused on generating the $10$ top values 
reached by the objective functions on configurations of the feature model at hand. 
Unlike what happened for top-$10$ configurations, the expected result is defined in a unique way as soon as the objective function is provided.

\begin{table}[b!]
\resizebox{\columnwidth}{!}{%
\begin{tabular}{l|c|c|c|}
\cline{2-4}
                                                       & success rate    & mean time  & max time   \\\hline
\multicolumn{1}{|l|}{\maxhs\ with $l=1$}           &  0.11           & 34.6s      & 154.6s     \\\hline
\multicolumn{1}{|l|}{\maxhs\ with $l=100$}         &  0.92           & 6.38s      & 73.0s      \\\hline
\multicolumn{1}{|l|}{\maxhs\ with $l=10000$}       &  0.99           & 4.73s      & 212s       \\\hline
\multicolumn{1}{|l|}{\maxhs\ with $l=1000000$}     &  1.0            & 4.45s      & 96.7s      \\\hline
\multicolumn{1}{|l|}{{\tt Sat4j} with $l=1$}           &  0.55           & 36.8s      & 542s      \\\hline
\multicolumn{1}{|l|}{{\tt Sat4j} with $l=100$}         &  0.81           & 12.7s      & 289s     \\\hline
\multicolumn{1}{|l|}{{\tt Sat4j} with $l=10000$}       &  0.83           & 13.2s      & 256s     \\\hline
\multicolumn{1}{|l|}{{\tt Sat4j} with $l=1000000$}     &  0.83           & 13.3s      & 212s     \\\hline
\end{tabular}
}
\caption{Top $10$ values, direct approach.}
\label{table:optVAL_CNF}
\end{table}

\paragraph{Direct Approach}

We ran the experiment using \satfourj\ and \maxhs. For enumerating the different values with \maxhs, we enumerate the optimal configuration until 10 different values are found.
The results are presented in \citetable{optVAL_CNF}.
Just like for computing $10$ top configurations, the choice of the bound $l$ turned out to have an impact on the performance of \satfourj\ and \maxhs\ for generating the $10$ top values. 
The approach using \maxhs\ failed for a value of $l=1$, which can be explained by the increase of optimal configurations when we decrease the value of $l$. 
In our methodology, for a CNF with $n$ variables, the value a configuration can take ranges between $0$ and $l \times n$.
With small values of $l$, the set of configurations has only a very limited space to distribute over and so for a given value the number of configurations increases. 
Since \maxhs\ has to enumerate configurations to find different values, the problem becomes closer to counting configurations for small values of $l$ and is exactly this problem for $l=0$. 
This explains the low success rate for \maxhs\ when $l$ is small.
\satfourj\ seems to struggle more for small values of $l$ and then stabilizes to a success rate of around 82\%.

\paragraph{Compilation-Based Approach}

We also ran the experiment using the compilation-based approach and the results are presented in \citetable{optVal_dDNNF}. The running time do not include the compilation time.
The compilation-based approach was able to solve in less than 8 seconds %(including the preprocessing time\pierrem{Compilation + loading?}) 
the top-$10$ values instances that have been considered, \ie for all the objective functions and all the feature models out of $221$ that have managed to compile.

\begin{table}[h]
\resizebox{\columnwidth}{!}{%
\begin{tabular}{l|c|c|c|}
\cline{2-4}
                                                       & success rate    & mean time  & max time  \\\hline
\multicolumn{1}{|l|}{\dDnnf\ with $l=1$}                &  0.995          & 0.33s      & 7.82s     \\\hline
\multicolumn{1}{|l|}{\dDnnf\ with $l=100$}              &  0.995          & 0.33s      & 7.56s     \\\hline
\multicolumn{1}{|l|}{\dDnnf\ with $l=10000$}            &  0.995          & 0.33s      & 7.67s     \\\hline
\multicolumn{1}{|l|}{\dDnnf\ with $l=1000000$}          &  0.995          & 0.32s      & 7.51s     \\\hline
\end{tabular}
}
\caption{Top $10$ values, compilation-based approach.}
\label{table:optVal_dDNNF}
\end{table}

\citefig{topval_maxhs_vs_jddnnf} shows how the direct approach using \maxhs\ compares with the compilation-based approach (compilation time included) for computing the top $10$ values.
In most case the compilation-based approach is faster than the direct approach and the compilation is amortized with one top $10$ value.

\begin{figure}[ht!]
\includegraphics[width=0.9\linewidth]{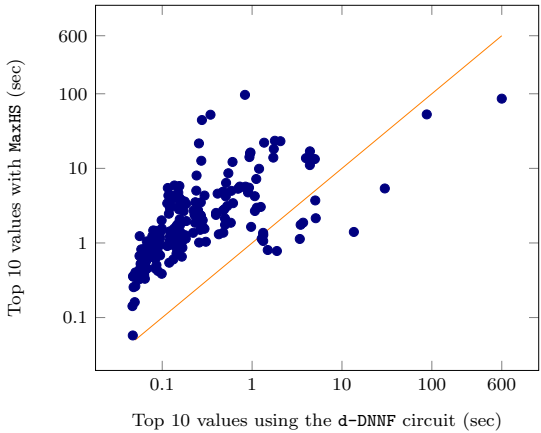}
\caption{Comparison of the top 10 values with \maxhs\ and the compilation-based approach with compilation time included. $l$ is set at 1000000.}
\label{fig:topval_maxhs_vs_jddnnf}
\end{figure}

\begin{formal}
The compilation-based approach was able to handle all the objective functions while the PB solver failed frequently and the \maxsat\ approach had a low success rate when the coefficients used in the objective function were smaller than 100.
\end{formal}

\section{Discussion and Threats to Validity}
\label{sec:discussion}

\paragraph{Performances}
As a rule of thumb, the compilation-based approach is typically advantageous when the off-line compilation phase finishes in a reasonable amount of time. 
Indeed, this implies that the size of the compiled form will remain ``small enough". 

\paragraph{Reasoning}
As sketched before, many reasoning operations become computationally easy when the input is a \dDnnf\ circuit. 
Especially, some optimization problems (that are in general {\sf NP}-hard) become tractable after compiling a FM into a \dDnnf\ circuit. 
Thus, determining a solution satisfying a given \dDnnf\ circuit and
maximizing the value of a given linear objective function (in the propositional variables) is feasible in polynomial time \cite{DarwicheMarquis04,Koricheetal16}.
That is, one can leverage such tractable operations to determine a preferred configuration of a FM and/or to compactly represent the set of all such preferred configurations when the preference relation can be modeled by a fully linear utility function. 

\paragraph{Validity Threats}
There exist a number of issues that might jeopardize the validity of our results. 
First, to ensure a correct implementation of \softName\ and avoid wrong or inconsistent results, we conducted a series of unit tests and checked the consistency of the results computed by the various algorithms. 
In particular, we used and tested against mature open-source software such as \satfourj\ and \dfour\ to further reduce possible inconsistencies.
Then, one might consider that the translation into the DIMACS format could be faulty.
For this translation, we relied on the FeatureIDE-library who had already been sanity checked for similar translations~\cite{Sundermann_VAMOS_20}.
There also exist threats that may affect the generalization of our results.
While we cannot claim that they can be transferred to all other feature models, we conducted our evaluation on a large set of FM (222) ranging from the automotive domain to software system.

\section{Conclusion}
\label{sec:conclusion}

In this paper, we compared a direct approach with a compilation-based approach relying on \dDnnf\ circuits on various analysis on feature models.
Through a number of experiments, we have shown that addressing a number of reasoning operations on FMs amounts to solving optimization problems and that \dDnnf\ are well-suited in this case.
Indeed, many queries of interest can be achieved in polynomial time from such representations, ensuring guaranteed response times.
Accordingly, the \dDnnf\ compilation of FMs looks very promising: our experiments have shown that the computational bottleneck lies only in the compilation step, so that when it succeeds many complex reasoning operations on FMs can be achieved efficiently. 
Despite this bottleneck, our experiments show that the compilation-based approach performs well to solve efficiently optimization problems over FMs and even some other reasoning tasks on the top.
In addition, we provided a Java-based tool support named \softName.
This tool is able to perform any classical reasoning operations on large configuration spaces (count configurations, enumerate or sample them, and even find the best one \emph{w.r.t.} an utility function) by taking indifferently as input a \cnf\ (direct approach) or a \dDnnf\ (compilation-based approach) representation of the FM.

As part of our future work, we plan to perform additional experiments, especially regarding attributed FM.
Of particular interest for further work is the language of \mddg\ representations, the extension of Decision-\Dnnf\ to constraints over variables having finite, yet (possibly) non-Boolean domains.
It would be interesting to define encodings associating FMs with constraint networks so as to exploit existing \mddg\ compilers \cite{Koricheetal16,Lagniezetal17}. 
It is expected that both the constraint network encoding and the \mddg\ compiled form of it will be much more compact than the 
corresponding \cnf\ formula and the associated Decision-\Dnnf\ circuit (respectively). This could lead to handling efficiently hard reasoning tasks
on much larger feature models, thus pushing forward the scalability of the approach.
\section*{Acknowledgements}

This work has been supported by the CPER DATA Commode project from the “Hauts-de-France” Region. It has also been partly supported by the PING/ACK project (ANR- 18-CE40-0011) and the KOALA project (ANR-19-CE25- 0003-01) from the French National Agency for Research.

\bibliography{main}

\end{document}